\documentclass[journal]{IEEEtran}
\normalsize
\ifCLASSINFOpdf
\else

\fi

\usepackage{mathtools}
\usepackage{amsmath}
\usepackage{amssymb}
\usepackage{amsmath}
\usepackage{cite}
\usepackage{multirow}
\usepackage{booktabs}
\usepackage{array}
\usepackage{siunitx}
\usepackage{graphicx}
\usepackage{graphicx}

\date{}
\usepackage{mdwmath}
\usepackage{blindtext}
\usepackage{eqparbox}

\begin{document}
\title{GOPA: Geometrical Optics Positioning Algorithm Using Spatial Color Coded LEDs\\ (Extended Version)}
\author{~Hamid Hosseinianfar, Ata Chizari, and~Jawad A.~Salehi,~\IEEEmembership{~Fellow,~IEEE}
\thanks{Hamid Hosseinianfar is with the Optical Multiuser/Multichannel Communications Lab. (OMCL), School of Engineering and Applied Science, University of Virginia. Ata Chizari is with the Biomedical Photonic Imaging group (BMPI), Faculty of Science and Technology, University of Twente. Jawad A. Salehi is with the Optical Networks Research Lab. (ONRL), Department of Electrical Engineering, Sharif University of Technology (e-mail: hh9af@virginia.edu, a.chizari@utwente.nl, jasalehi@sharif.edu).}

}
\maketitle
\begin{abstract}
	In this paper, we propose an accurate visible light indoor localization system for a smartphone using the existing commercial light-emitting diode (LED) luminaries. The proposed technique, called geometrical optics positioning algorithm (GOPA), uses spatial color code landmarks alongside angle of arrival (AOA)-based geometrical algorithm on smartphones to locate the device, and reserves LED's time-frequency domain modulation to increase the throughput of the visible light network broadcast from the same luminaries infrastructure. GOPA algorithm is developed with practical considerations such as flexible hand gesture and handshake, and it enables both positioning robustness and on-device processing. By introducing the idea of virtual plane, field of view (FOV) issue of AOA-based positioning systems is addressed in GOPA. The embedded accelerometer and front-facing camera of the smartphone are used at the receiver side to measure the smartphone inclination and acquire the image. Experimental results show robust two-dimensional ($2$-D) and three-dimensional ($3$-D) positioning. The experimental mean positioning error for $2$-D positioning is $0.54$ cm, in case one ignoring the tilt. The experimental mean positioning errors for $3$-D positioning are respectively $1.24$ cm, $1.85$ cm, and $6.02$ cm for ideal non-tilted and non-oriented, non-tilted but orientated, and both tilted and orientated scenarios.

\end{abstract}


%
\IEEEpeerreviewmaketitle
\section{Introduction}
%
%
%
%

\IEEEPARstart{T}{he} ever-increasing demands for location-based services in a broad range of applications such as product tracking, indoor navigation services, and location-based advertisements have raised the necessity of indoor positioning. 
In the retail sector, remarkable revenue enhancement is estimated, if to-the-point advertisements and smartphone indoor navigation are employed \cite{6685754,luo2017indoor,ho2017visible,zhuang2018survey}. 

Various indoor positioning systems have been proposed over the last decade. The radio-frequency (RF) techniques take advantage of extremely low cost and broad infrastructures. Nevertheless, these techniques have poor resolution, are sensitive to the background of the environment and miss the orientation information. Hence, they are not utilized for most commercial applications such as shelf-level advertising \cite{chintalapudi2010indoor,4343996}. Moreover, the radio-frequency operation is forbidden in hospitals, near airports and also in military fields due to the disturbance of other equipment.

Visible light positioning (VLP) systems have been recently proposed for indoor environments due to integrability with exponentially deployed light-emitting diode (LED) luminaries \cite{steendam2017theoretical}. These systems are considered to be superior due to their high positioning accuracy \cite{zhang2013comparison}. Moreover, their complementary system, namely visible light communication (VLC), is the apt approach for indoor data access providing higher multiuser capacity \cite{noshad2013can,hosseinianfar2017positioning,chizari2017visible,zhang2018anticipatory}. Exploiting the ubiquitous presence of smartphones and their powerful built-in sensors such as front-facing cameras and accelerometers, development of visible light indoor positioning algorithm using unmodified smartphones is the subject of ongoing research in this area \cite{Kuo2014,yasir2014indoor,6950776,li2014epsilon, li2018vlc}.

Since the angle of arrival (AOA) is the most promising method for VLP \cite{6685759, Kuo2014}, in this paper we propose geometrical optics positioning algorithm (GOPA), a comprehensive smartphone-based indoor positioning approach in which spatial color-coded landmarks are used as identifiers of positioning reference points. The algorithm utilizes simple image processing to extract color code in the images captured by the front-facing camera. Therefore, by searching the extracted landmarks from the pre-stored database, we obtain the global position of the receiver. Highly accurate local positioning is performed using two reference points from the captured landmark in the geometrical optics algorithm. Moreover, a novel virtual plane idea embedded in the geometrical optics algorithm addresses both the challenges of tilt and the limitation of the field of view (FOV) in realistic applications.
  The simple local positioning and landmark code extraction procedures reduce the complexity of processing and take advantage of smartphone's operating system to perform all processing.

The rest of the paper is organized as follows. Comparison of GOPA with other VLC-based positioning systems
is presented in Section II. Section III describes notations and overviews the whole system.
The local positioning concept containing $2$-D and $3$-D positioning, followed by azimuth and tilt considerations will be investigated in Section IV. The global positioning algorithm is 
studied in Section V. Section VI provides experimental results and discussions, and Section VII concludes the paper.

\section{Related Works}
The main goal of most indoor positioning applications is to estimate the position within a fixed structure of transmitters as the reference points. Hence, the position error, i.e., the Euclidean distance between real location of the target and the measured point, is the main metric considered in the evaluation of the proposed indoor positioning systems. In this scope, we narrow our investigation down to visible light indoor positioning systems, which show promising accuracy in comparison to RF-based and infrared (IR)-based systems \cite{zhang2013comparison,Kuo2014}.

\begin{table*}[t]
	\begin{center}
		\renewcommand{\arraystretch}{1}
		\caption{A comparison to prior VLP systems on smartphone and conventional optical indoor positioning systems.}
		\label{table_Comp}
		\footnotesize
		\begin{tabular}{|c|c|c|c|c|c|c|c|l|}
			\hline
			\multirow{4}{*}{\textbf{\begin{tabular}[c]{@{}c@{}}Reference \\ No.\end{tabular}}} & \multirow{4}{*}{\textbf{\begin{tabular}[c]{@{}c@{}}Positioning\\ Algorithm\end{tabular}}} & \multirow{4}{*}{\textbf{\begin{tabular}[c]{@{}c@{}}Mean  position\\ Error (mm)\end{tabular}}} & \multirow{4}{*}{\textbf{\begin{tabular}[c]{@{}c@{}}Space\\ Dimension\end{tabular}}} & \multirow{4}{*}{\textbf{\begin{tabular}[c]{@{}c@{}}Device\\ Orientation\end{tabular}}} & \multirow{4}{*}{\textbf{\begin{tabular}[c]{@{}c@{}}Tilt\\ Consideration\end{tabular}}} & \multicolumn{3}{c|}{\textbf{Required Knowledge}}                                                                                                  \\ \cline{7-9} 
			&                                                                                           &                                                                                               &                                                                                     &                                                                                        &                                                                                        & \multirow{3}{*}{\textbf{Channel}}    & \multicolumn{2}{c|}{\multirow{3}{*}{\textbf{\begin{tabular}[c]{@{}c@{}}Transmitter \\ Term\end{tabular}}}} \\
			&                                                                                           &                                                                                               &                                                                                     &                                                                                        &                                                                                        &                                      & \multicolumn{2}{c|}{}                                                                                      \\
			&                                                                                           &                                                                                               &                                                                                     &                                                                                        &                                                                                        &                                      & \multicolumn{2}{c|}{}                                                                                      \\ \hline \hline
			\cite{6131130}                        & TDOA                                                                                      & $1.8$ (Sim.)                                                                                    & $2$-D                                                                                  & Not supported                                                                          & No                                                                                     & Yes                                  & \multicolumn{2}{c|}{\begin{tabular}[c]{@{}c@{}}Identifier\\  synchronization\end{tabular}}                 \\ \hline
			\cite{zhou2012indoor}                 & RSS                                                                                       & $0.5$ (Sim.)                                                                                    & $2$-D/$3$-D                                                                               & Not supported                                                                          & No                                                                                     & Yes                                  & \multicolumn{2}{c|}{Source power}                                                                          \\ \hline
						\cite{hosseinianfar2018analysis}                 & Fingerprinting                                                                                       & $50$ (Sim.)                                                                                    & $2$-D                                                                               & Not supported                                                                          & No                                                                                     & Yes                                  & \multicolumn{2}{c|}{---------}                                                                          \\ \hline
			\cite{6950776}                        & RSS/ DTMF                                                                                 & $16$ (Exp.)                                                                                     & $2$-D                                                                                  & Not supported                                                                          & No                                                                                     & Yes                                  & \multicolumn{2}{c|}{---------}                                                                                 \\ \hline
			\cite{5381102}                        & AOA                                                                                       & $50$ (Exp.)                                                                                     & $3$-D                                                                                  & Supported                                                                              & Yes                                                                                    & No                                   & \multicolumn{2}{c|}{---------}                                                                                 \\ \hline
			\cite{6823667}                        & RSS/AOA                                                                                   & $30$ (Exp.)                                                                                      & $2$-D/$3$-D                                                                               & Not Supported                                                                          & No                                                                                     & Yes                                  & \multicolumn{2}{c|}{Source power}                                                                          \\ \hline
			\cite{li2014epsilon}                  & RSS                                                                                       & $400$ (Exp.)                                                                                     & $3$-D                                                                                  & Supported                                                                              & Yes                                                                                    & Yes                                  & \multicolumn{2}{c|}{Source power}                                                                          \\ \hline
			\cite{Kuo2014}                        & AOA                                                                                       & $290$ (Exp.)                                                                                     & $3$-D                                                                                  & Supported                                                                              & Yes                                                                                    & No                                   & \multicolumn{2}{c|}{---------}                                                                                 \\ \hline
			\cite{yasir2014indoor}                & RSS                                                                                       & $250$ (Exp.)                                                                                     & $2$-D/$3$-D                                                                               & Supported                                                                              & Yes                                                                              & Yes                                  & \multicolumn{2}{c|}{Source power}                                                                          \\ \hline
			GOPA                              & AOA                                                                                       & $60$ (Exp.)                                                                                      & $2$-D/$3$-D                                                                                  & Supported                                                                              & Yes                                                                              & No                                   & \multicolumn{2}{c|}{---------}                                                                                 \\ \hline
		\end{tabular}
	\end{center}
\end{table*}

Table \ref{table_Comp} compares recent visible light positioning systems that have employed the aforementioned methods. In this comparison, systems are evaluated based on the employed technique, mean position error, the space dimension, and practical terms consideration. The time difference of arrival (TDOA)-based indoor positioning, which estimates receiver position based on the difference of signal's arrival time from different reference points, shows $1.8$ \si{mm} accuracy in simulation \cite{6131130}. However, this simulation is based on a simple line of sight (LOS) channel, assuming the receiver axis is fixed. In addition, due to the short propagation time of light, especially in a short-distance indoor environment, the accuracy of these algorithms degrades significantly due to synchronization error. 

Received signal strength (RSS)-based method is proposed as a potential approach in VLC-based indoor positioning systems. In  \cite{zhou2012indoor}, an accuracy of around $0.5$ mm is demonstrated in simulation, neglecting reflection components of the emitted patterns.
However, this accuracy is degraded when practical considerations such as receiver axis tilt variation are taken into account, as it is demonstrated in \cite{6880333} in terms of Cramer-Rao bound of accuracy in the scale of Centimeter, neglecting multipath effects.
 Moreover, multipath effects cause extra positioning errors and algorithm complexity in realistic systems, as the accuracy of $25$ \si{cm} is obtained in \cite{yasir2014indoor}. While RSS technique considers the multipath signal as noise, a new fingerprinting technique has been developed in \cite{7996815,hosseinianfar2018analysis}, which uses the characteristics of the optical channel impulse response in order to lacate users.
Emerging from the high spatial resolution nature of light, the angle of arrival technique is considered as the most promising method for new VLP systems since it resolves the destructive multipath effects on the positioning accuracy \cite{6685759}. This method is robust against channel background changes; therefore, it shows more accurate positioning in comparison to other methods.

The proposed system demonstrates an accuracy of $0.5$ \si{cm} for $2$-D and $1.24$ \si{cm} for $3$-D positioning in practice.  Given the worst smartphone's inclination, the positioning robustness is preserved with a mean error about $6.02$ \si{cm}, which makes this system the most comprehensive, highly accurate indoor positioning system reported to date.



\section{Notations and System Overview}
\subsection{Notations}
\begin{figure}[!t]
	\centering
	\includegraphics[width=2.9in]{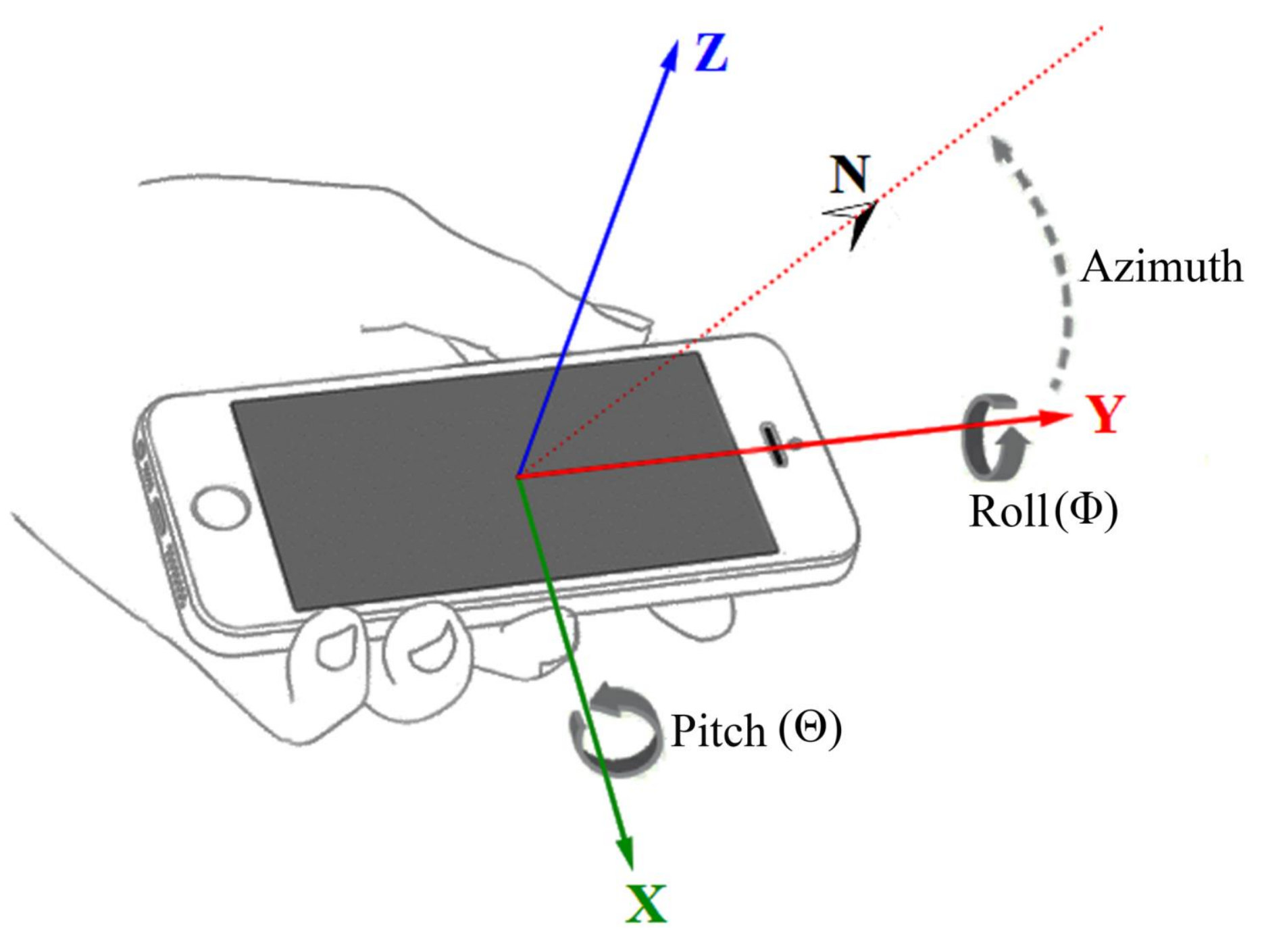} 
	\caption{Demonstration of roll, pitch and azimuth.}
	\label{fig:Notation}
\end{figure}

In this subsection, we clarify the meaning of terms used throughout the paper. Fig. \ref{fig:Notation} illustrates the smartphone inclination terms. The rotation of the camera on the horizontal plane ($xy$ plane) is known as the azimuth ($\Psi $)  or orientation, and its rotations around $x$ and $y$ axes are denoted as pitch($\Theta $) and roll ($\Phi $), respectively. Roll and pitch inclinations are denoted tilt as well. The spatial color-coded landmark is defined as a set of color LEDs that constitute a unique spatial identifier of a positioning cell (see Fig. \ref{subfigure1}-(b)). 
In this work, we consider local positioning as the calculation of the positioning parameters inside a landmark's cell. In addition, finding the cell in which the smartphone is located is called global positioning. Moreover, the Cartesian coordinate system axes of the camera and the real world are called camera coordinate system axes and room coordinate system axes, respectively. Lastly, a triangle with vertices $A$, $B$, and $C$ is denoted by $\triangle \overline{ABC}$.
\subsection{System Overview}
\begin{figure}[t]
	\centering
	\includegraphics[width=3.55in]{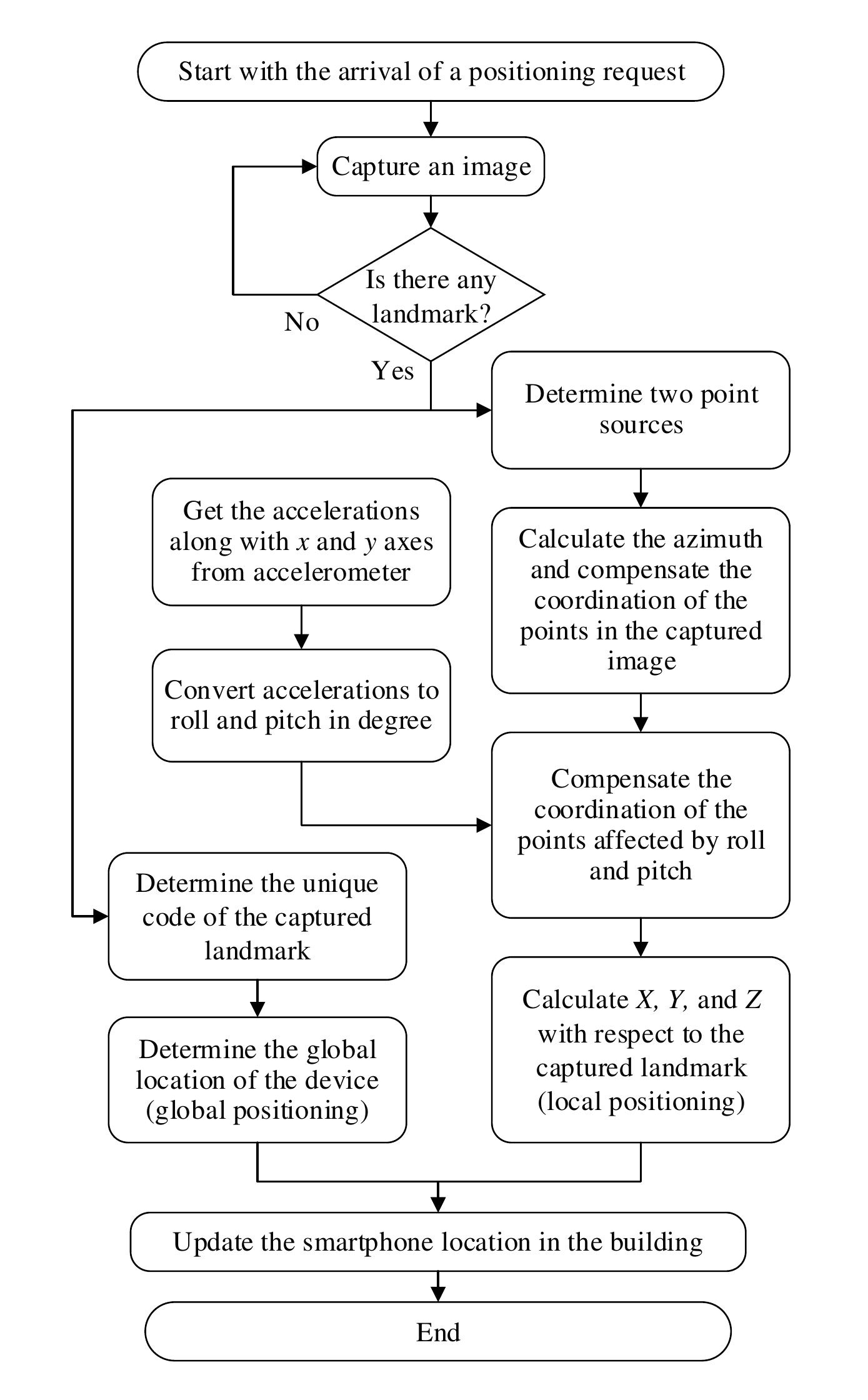} 
	\caption{Flowchart of the proposed algorithm.}
	\label{Diagram}
\end{figure} 
The proposed indoor positioning system consists of colored LEDs landmarks and a smartphone equipped with an accelerometer and a front-facing camera. Each landmark includes a set of color-coded LEDs, namely, red, green, and blue, which create a spatial identifier of the corresponding landmark. Fig. \ref{Diagram} illustrates the flowchart of the proposed algorithm. In this algorithm, the front-facing camera of the smartphone continuously takes pictures upon the positioning request. The algorithm then checks whether there is any landmark in the image. In case a landmark exists, a simple image processing algorithm is employed in order to extract the landmark's unique color code and derive coordinates of two reference points in the landmark. In order to find the distance of the smartphone from the captured landmark, the local positioning algorithm checks whether the smartphone has a non-zero azimuth and considers this parameter in the positioning calculations.
Section \ref{sec:algorithm with azimuth consideration} discusses the details of azimuth consideration. Furthermore, the values of the smartphone's roll and pitch measured by the accelerometer sensor are taken into account in the positioning algorithm, which will be discussed in Section \ref{sec:Tilt Consideration}. After calculation of the azimuth and tilt compensation vectors, we apply our geometrical optics-based algorithm to find the position of the smartphone within the positioning cell. Both the $2$-D and $3$-D localization algorithms are discussed thoroughly in Sections \ref{sec:2D localization with zero roll and pitch} and \ref{Sec: 3D localization with zero roll and pitch}, respectively. Ultimately, the general coordinates could be evaluated using a lookup table of landmarks identifiers saved in the smartphone which is discussed in Section \ref{sec:global positioning}.
\section{Local positioning}
\label{sec:Local positioning}
In the local positioning, we propose a novel method which takes advantage of requiring only one point source for a highly accurate $2$-D positioning. In addition, for an accurate $3$-D positioning, our method only needs two point sources, that are fixed at the same height from the ground. While in counterpart AOA-based technique, namely, the triangulation method, two and three reference points are needed to perform $2$-D and $3$-D positioning, respectively \cite{zhang2013comparison}. In what follows we discuss different aspects of our proposed algorithm.
\subsection{$2$-D Localization with Zero Roll and Pitch}
\label{sec:2D localization with zero roll and pitch}
\begin{figure}[t]
	\centering
	\includegraphics[width=3.2in]{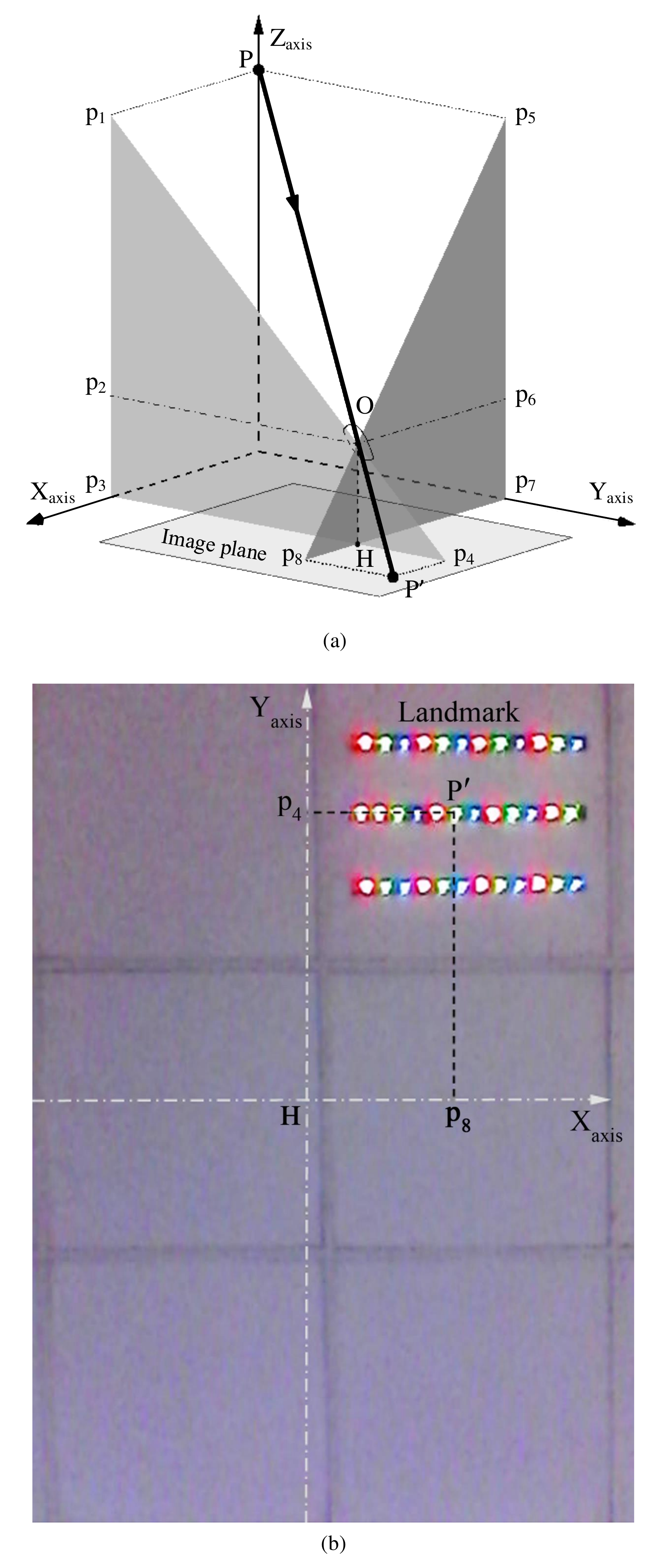} 
	\caption{Concept of $2$-D positioning: (a) Magnified geometric scheme. (b) Corresponding captured image by the front-facing camera.}
	\label{subfigure1}
\end{figure}

In this section, we explain the proposed simple $2$-D positioning algorithm, assuming the camera has zero tilt and zero azimuth, i.e., both the room's and camera's Cartesian coordinate systems are the same. Given that rays passing through the center of the biconvex lens exit in the same direction, the projection of the LEDs on the image plane (detector's area) is collinear with the LEDs and the center of biconvex lens \cite{Kuo2014}. Fig. \ref{subfigure1}-(a) illustrates the geometric scheme of the proposed algorithm. The projection of the lens's center on the image plane, $H$, is the origin of the coordination system. Assuming the point source $P$ is in the camera's FOV, the projection of $P$ on the camera's image plane, $P'$, has coordinates $(P_8, P_4)$ on the $xy$ plane of the camera's coordinate system. The desired positioning parameters are the horizontal distance of $P$ and $O$, namely  $X=\overline{P_{7} H}$ and $Y=\overline{P_{3} H}$. In addition, $Z=\overline{P_{1} P_{2} }=\overline{P_{5} P_{6} }$ is the height of the smartphone with respect to $P$, which is known in $2$-D positioning. The similarity of triangles $\triangle \overline{P_{5} P_{6}O}$ and $\triangle \overline{OHP_{8}}$ yields

\begin{equation}
{\frac{Z}{\overline{OH}} }=\frac{X}{\overline{HP_{8} }}.
\label{2Deq1}
\end{equation}

In the same way, considering the similarity of $\triangle \overline{P_{1}P_{2}O}$ and $\triangle \overline{OHP_{4}}$, we obtain

\begin{equation} 
\label{2Deq2} 
{\frac{Z}{\overline{OH}} }=\frac{{Y}}{\overline{HP_{4} }},  
\end{equation} 
where $\overline{OH}=Z_{c}$ is the focal length of the camera. All the distances in (\ref{2Deq1}) and (\ref{2Deq2}) are in centimeters.

Fig. \ref{subfigure1}-(b) illustrates the captured image corresponding to the image plane in Fig. \ref{subfigure1}-(a). Considering that all the distances in the captured image are measured in pixels, the $P'$ distance components from $H$, i.e, $(\overline{HP_{8}} ,\overline{HP_{4}})$ can be defined as
\begin{align} 
\label{2Deq3} 
\overline{HP_{8}}= UP_{x}, \nonumber \\
\overline{HP_{4} }=UP_{y},
\end{align} 
where $U$ is a conversion constant in \si{cm/pixel}, $P_{x}$ and $P_{y}$ are the $\overline{P'H}$ components in \si{pixels}. In addition, 
$Z_{c} $ and $U$ are constants that depend on camera parameters, and they must be calculated for each camera once. This calibration mechanism is discussed further in Section \ref{sec:Experimental results and discussion}. Ultimately, substituting (\ref{2Deq3}) in (\ref{2Deq1}) and (\ref{2Deq2}), the desired parameters can be obtained as
\begin{align} \label{2Deq4} 
X=\frac{Z}{Z_{c}} P_{x}U, \nonumber \\
Y=\frac{Z}{Z_{c}} P_{y}U,  
\end{align}
which verifies the feasibility of $2$-D positioning using just one point source.
\subsection{$3$-D Localization with Zero Roll and Pitch}
\label{Sec: 3D localization with zero roll and pitch}
\begin{figure}[t]
	\centering
	\includegraphics[width=3.2in]{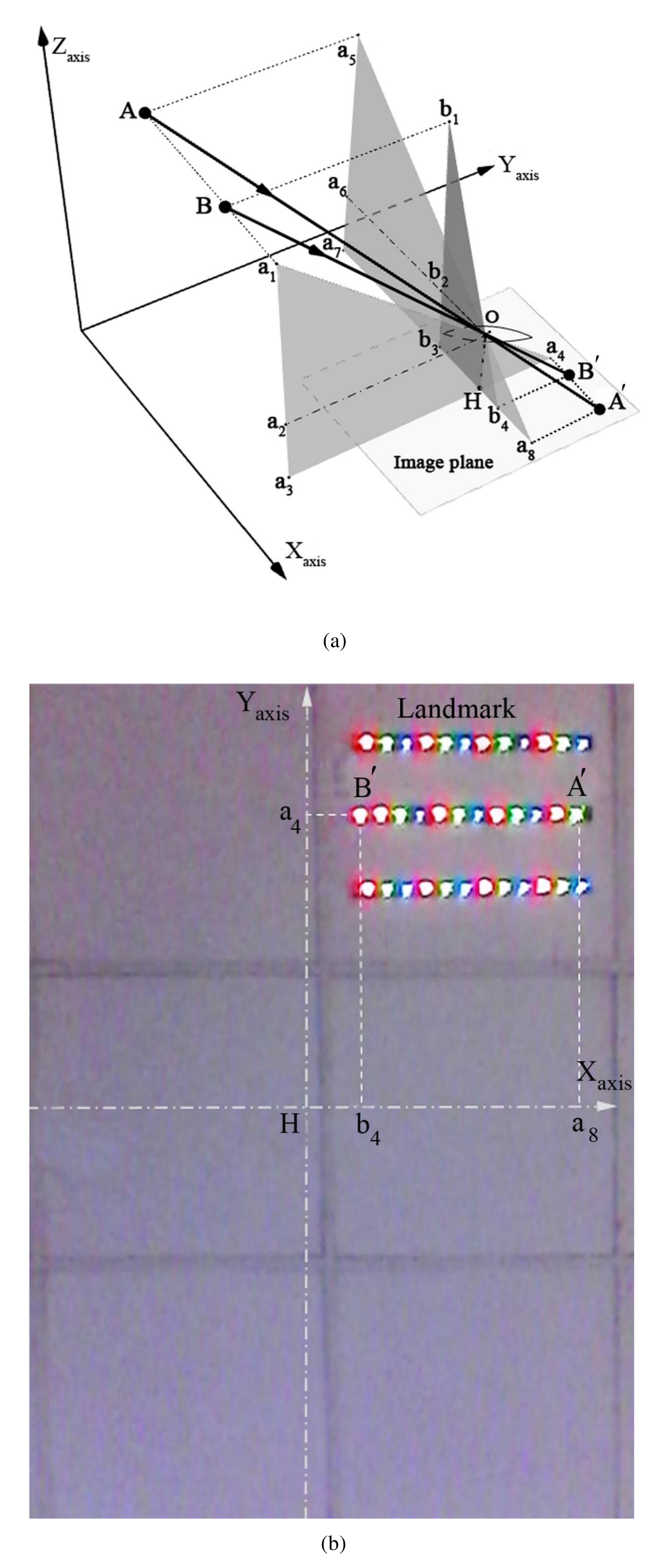} 
	\caption{Concept of $3$-D positioning: (a) Magnified geometric scheme. (b) Corresponding captured image by the front-facing camera.}
	\label{subfigure2}
\end{figure}

In order to perform $3$-D positioning, extraction of two point sources' coordinates in the captured image of a landmark is required. Let us first assume that the camera's azimuth is zero. Fig. \ref{subfigure2}-(a) shows the geometric scheme of $3$-D positioning. Points $A$ and $B$ illustrate the extracted point sources with the same $y$ and $z$ components, i.e., $\overline{AB}$ is parallel to the $x$ axis. Similar to Section \ref{sec:2D localization with zero roll and pitch}, the projected points on the image plane, namely $A'$ and $B'$, have the same $y$ but different $x$ coordinates. Furthermore, considering zero the roll and pitch for the camera, the $z$ axis is orthogonal to the image plane. Accordingly, the room coordinate system is shown in Fig. \ref{subfigure2}-(a) is similar to the camera coordinate system, shown in Fig. \ref{subfigure2}-(b). The desired parameters of $3$-D positioning are the components of $\overline{AO}$ distance, namely $X=\overline{a_{7} H}$, $Y=\overline{a_{3} H}$ and $Z=\overline{a_{5} a_{6}}=\overline{b_{1} b_{2}}=\overline{a_{1} a_{2}}$. In this scenario, there are three triangles' similarities, shown in Fig. \ref{subfigure2}-(a). The subsequent distance relations can be derived as follows.

\textit{1)} Since $\triangle \overline{a_{5} a_{6}O}$ and $\triangle \overline{OHa_{8}}$ are similar, it can be seen that
\begin{align}
\frac{Z}{Z_{c} } &=\frac{X}{p_{x_{a} } U},
\label{3Deq1}
\end{align}
where $Z_{c} $ and $U$ are the conversion constants introduced in the $2$-D positioning subsection, and $X=\overline{a_{7} H}$. Furthermore, $p_{x_{a} } $ is the number of pixels between $H$ and $a_{8} $ in the captured image. 

\textit{2)} The similarity between $\triangle \overline{b_{1} b_{2}O}$ and $\triangle \overline{OHb_{4}}$ yields
\begin{align}
\frac{Z}{Z_{c} } &=\frac{X-L}{p_{x_{b} } U},
\label{3Deq2}
\end{align}
where $L=\overline{a_{7} b_{3} }$ is the known fixed distance between the point sources $A$ and $B$, which are located at the same landmark. Similarly, $p_{x_{b}} $ is the number of pixels between $H$ and $b_{4} $ in the captured image.

\textit{3)} Considering the similarity of $\triangle \overline{a_{1} a_{2}O}$ and $\triangle \overline{OHa_{4} }$, we obtain
\begin{align}
\frac{Z}{Z_{c} } &=\frac{Y}{p_{y} U},
\label{3Deq4}
\end{align}
where $Y=\overline{a_{3} H}$ and  $p_{y}$ is the number of pixels between $H$ and $a_{4} $ in the captured image. Solving the system of linear equations including (\ref{3Deq1}), (\ref{3Deq2}) and (\ref{3Deq4}), the desired parameters $X$, $Y$, and $Z$ can be calculated as
\begin{align}
X =& L\frac{p_{x_{a} } }{p_{x_{a}} -p_{x_{b}}},  \nonumber \\
Y =& L\frac{p_{y} }{p_{x_{a}} -p_{x_{b}}},  \\
Z =& LZ_{c} \frac{1}{U(p_{x_{a}} -p_{x_{b}})}.  \nonumber
\label{3Deq5} 
\end{align}


\subsection{Azimuth Consideration}
\label{sec:algorithm with azimuth consideration}
In the above $2$-D and $3$-D positioning algorithms, it is assumed that the camera has zero azimuth and tilt. In other words, the camera and room coordinate systems are the same. In this part, the positioning algorithms are generalized for different values of the camera's azimuth, when the camera coordinate system is oriented around the $z$ axis with respect to the room coordinate system. For this purpose, the coordinates of two point sources, i.e., LEDs, in the captured image have to be extracted. 
\begin{figure}[!t]
	\centering
	\includegraphics[width=3.2in]{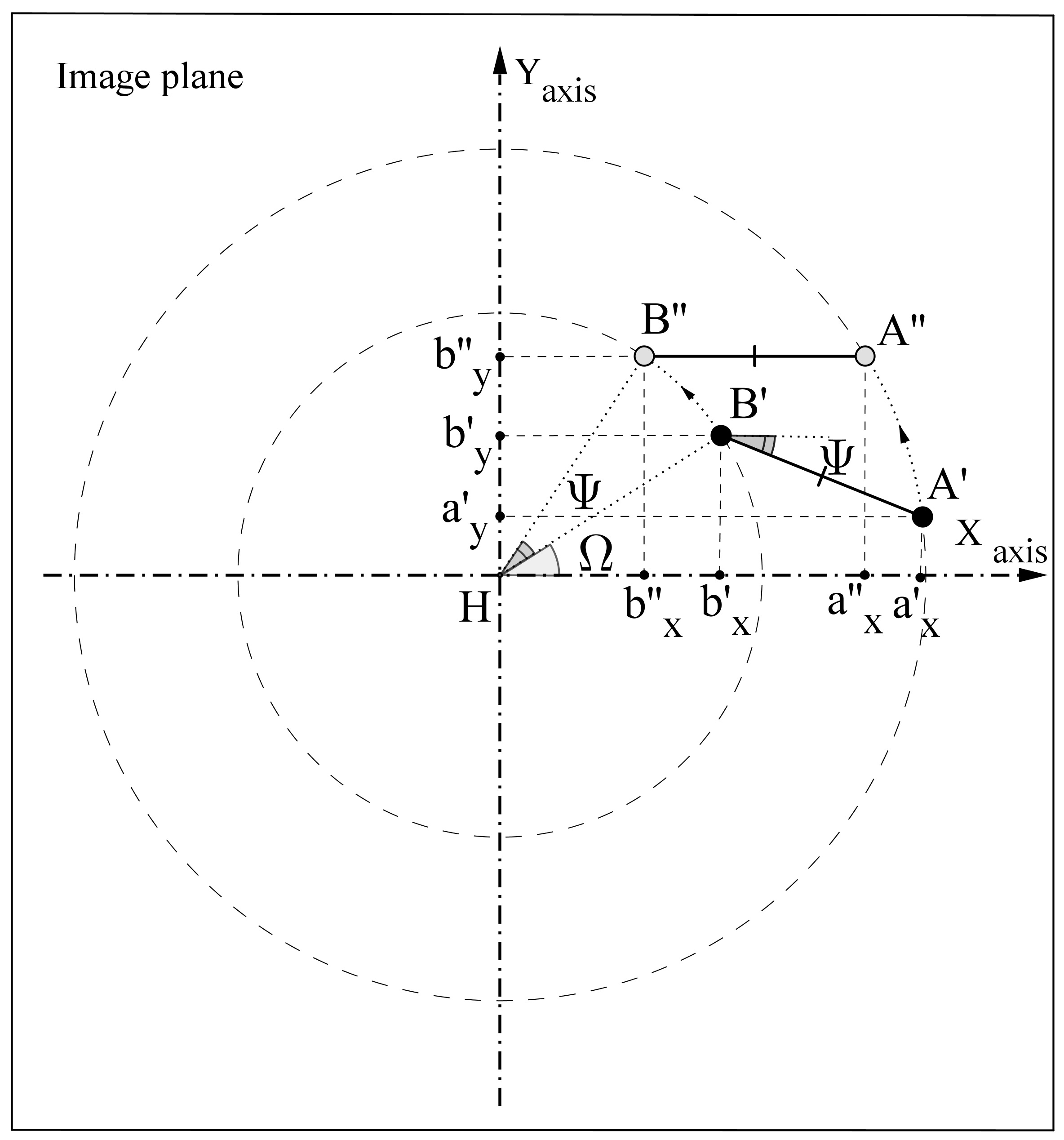} 
	\caption{Demonstration of the point sources location in the captured image affected by $\Psi $ degree azimuth.}
	\label{fig:Azimuth scheme}
\end{figure}
Fig. \ref{fig:Azimuth scheme} demonstrates the azimuth calculation scheme. The projected image of the point sources $A$ and $B$ on the image plane are $A'$ and $B'$, respectively, where $\overline{AB}$ is parallel to the $x$ axis of the room coordinate system. Considering that the camera is clockwise rotated by $\Psi$ degree, the segment $\overline{A'B'}$  and the $x$ axis of the camera coordinate system have $\Psi $ degree angle with respect to each other. Therefore, it is necessary to apply a $-\Psi $ degree rotation on $A'$ and $B'$ before employing the proposed $2$-D and $3$-D algorithms. Considering $A'=(a_{x}^{'},a_{y}^{'})$ and $B'=(b_{x}^{'},b_{y}^{'})$, the rotation value,  $\Psi $, can be obtained as
\begin{equation} \label{azimEqu1} 
\Psi =\tan ^{-1} (\frac{b_{y}^{'}-a_{y}^{'}}{a_{x}^{'} -b_{x}^{'}}).
\end{equation}

Accordingly, the corresponding rotation of $B'$, i.e., $B''=(b_{x}^{''},b_{y}^{''})$ can be calculated as
\begin{align} \label{azimEqu2} 
& r_{1}=\sqrt{(b_{x}^{'} -H_{x})^{2} +(b_{y}^{'} -H_{y} )^{2}},  \nonumber \\
&\Omega=\tan ^{-1} \frac{b_{y}^{'} -H_{y}}{b_{x}^{'} -H_{x}},\nonumber\\
&(b_{x}^{''},b_{y}^{''})=(r_{1} \cos (\Psi +\Omega ),r_{1} \sin (\Psi +\Omega )),
\end{align} 
where $r_{1}$ and $\Omega$ are the radius of rotation around $H=(H_{x}, H_{y})$ and the initial angle, respectively. The corresponding rotation of $A'$, i.e., $A''=(a_{x}^{''}, a_{y}^{''})$, can also be calculated as
\begin{align} \label{azimEqu4} 
& r_{2}=\sqrt{(b_{y}^{'} -a_{y}^{'} )^{2} +(a_{x}^{'} -b_{x}^{'} )^{2} },\nonumber\\
&(a_{x}^{''},a_{y}^{''})=(b_{x}^{''} +r_{2},b_{y}^{''}),
\end{align}
in which $r_2=\overline{A'B'}$. Now $\Psi $ is reported as the azimuth of the positioning device. In addition, $A^{''}$ and $B^{''}$ on the image plane are employed in the proposed positioning algorithms as the projection of the point sources $A$ and $B$, respectively.
		
\subsection{Tilt Consideration}
\label{sec:Tilt Consideration}
The other highly significant challenge of smartphone-based VLP systems is related to the tilt of the smartphone. In this part, we address this challenge through the virtual plane approach. In this regard, we employ an accelerometer, a built-in sensor that is available in most smartphones, in order to evaluate the smartphone's roll and pitch parameters used in virtual plane calculations.

\begin{figure}[!t]
	\centering
	\includegraphics[width=3.2in]{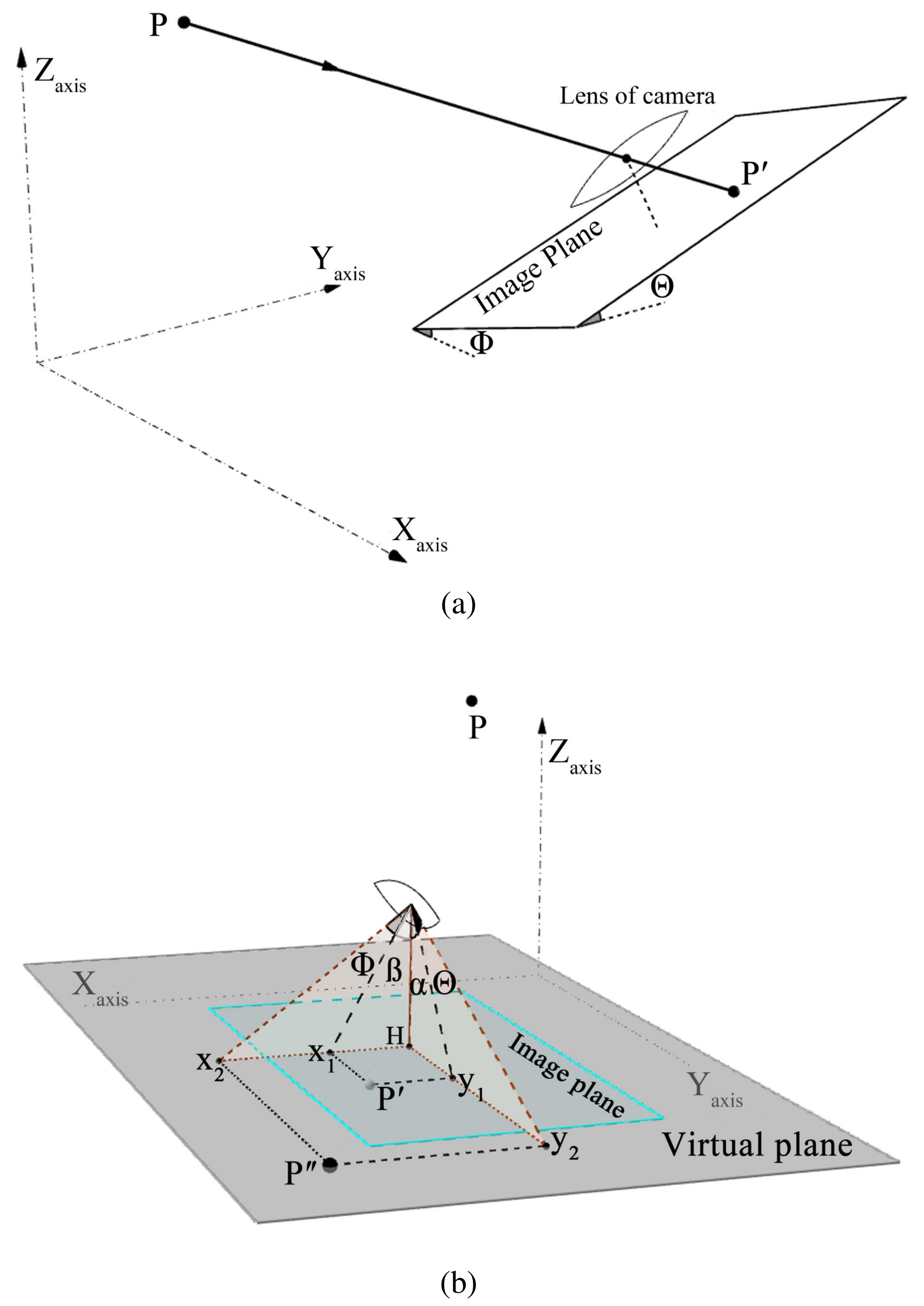} 
	\caption{Demonstration of tilt compensation: (a) Real gesture. (b) Compensated point $P^{''} $ on the virtual plane.}
	\label{subfigure3}
\end{figure}

In practice, there is always an inevitable tilt applied to the camera when users hold their smartphone. This leads the point sources to be projected on different points on the image plane and it makes this situation different from the situation with zero tilt, i.e., when the camera is absolutely horizontal. Fig. \ref{subfigure3}-(a) illustrates a realistic inclination of a smartphone with $\Phi $ degree roll and $\Theta $ degree pitch. The point source $P$ is in the FOV of the camera, hence its projected image, $P^{'} $, is located on the image plane. Fig. \ref{subfigure3}-(b) illustrates the locations of $P^{'}$ on the image plane, where $\alpha$ and $\beta$ are the angles that the $P$ beam vector makes with the $xz$ and $yz$ planes in the non-zero tilt situation, respectively. Now, assume the camera tilt is zero at the same location. In this case, the image of $P$ is projected at $P^{''}$ on the image plane and the aforementioned angles change to $(\alpha+\Theta)$ and $(\beta+\phi)$, respectively. Therefore, in order to address the tilt problem in the proposed algorithm, the map vector from $P'$ to $P''$ should be derived. Then, accurate positioning is possible, applying $P''$ as the projection of point source $P$ in the positioning algorithms. As it is shown in Fig. \ref{subfigure3}-(b),  $P^{''} $ might be located beyond the image plane due to the applied tilt compensation. That is why we call this approach a \textit{virtual plane}, owing to its virtually broadened image plane. We name the point $P^{''}$ as a virtual projected point. Considering $P'=(x_1, y_1)$ coordinates on the image plane, $\alpha$ and $\beta$ are obtained as
\begin{align}\label{tiltequ1} 
\alpha=\tan ^{-1} (U{(y_{1} -H_{y} )}/{\overline{OH}} ),\nonumber\\
\beta=\tan ^{-1} (U{(x_{1} -H_{x} )}/{\overline{OH}} ),
\end{align}
where $\overline{OH}=Z_{c}$ and $U$ are constant parameters in \si{cm} introduced earlier. $H=(H_{x},H_{y})$ is the center point of the image plane in \si{pixels} and the $P^{''}=(x_{2} ,y_{2})$ coordinates can be calculated as
\begin{align} 
{y_{2} =({Z_{c} }/{U}) \tan (\alpha +\Theta )+H_{y} },\nonumber \\
{x_{2} =({Z_{c} }/{U}) \tan (\beta +\Phi )+H_{x} }, 
\label{tiltequ2}
\end{align}
in which the roll and pitch angles are extracted using the accelerometer of the smartphone as follows:
\begin{align}  \label{tiltequ3} 
{\Phi =\sin ^{-1} ({\varphi }/{9.8} )},\nonumber\\
{\Theta =\sin ^{-1} ({\theta }/{9.8} )},
\end{align} 
where $\varphi$ and $\theta $ are the smartphone acceleration along $y$ and $x$ axes in \si{{m}/{s^{2}}} respectively, and $9.8$ is the gravity acceleration. Eventually, $P^{''}$ is a valid point for the proposed $2$-D positioning algorithm. In order to perform $3$-D positioning, it is necessary to compensate the coordinates of the projected image of two point sources of the same landmark.

The virtual plane approach addresses the tilt issues reported in most AOA-based methods \cite{Kuo2014}, where it gives flexibility to the user's hand gesture and guarantees localization accuracy even if the user's handshakes. Moreover, tilt compensation broadens the system FOV, which is the most critical issue in AOA-based methods \cite{zhang2013comparison}. In other words, by intentionally applying the tilt, the point source $P$ appears in the FOV of the positioning system.

\section{Global Positioning}
\label{sec:global positioning}
In the local positioning, we considered each landmark as a  positioning cell in which the user's location is measured based on the center of the landmark. In order to cover a larger area, such as a library building, shopping center, etc., each cell should have a unique identifier. 

In this section, we discuss spatial color code design and introduce the corresponding constraints that should be taken into account.  In the proposed indoor positioning system, the spatial arrangement of color LEDs in the landmark creates the cell identifier, and the other parameters, such as illumination level and time-frequency domain are reserved for VLC. As far as illumination is concerned, the color codes should be balanced. This means that the number of different color LEDs, namely, red, green and blue, should be equal, which leads to almost white illumination from landmarks. 

The other point on the code design has to do with the azimuth consideration in local positioning. The identifier codes and their orientations should remain unique compared to other subset members as well as their transposed and mirrored versions. For example, in a symmetric landmark, such as a square, the $90^{\circ}, 180^{\circ}$, and $270^{\circ}$ rotations of each code matrix are also unique. One of the most applicable approaches for simplifying orientation constraints is using a specific header in the code for absolute compass direction. For example, a red row of LEDs in the north side of a square landmark or a red radius of LEDs toward the north in a circle landmark handles orientation concerns in the code design. However, in designing the rest of the code matrices, the same red row in the other sides of the square or the same radius in the circle must be avoided.

Now, in order to evaluate the scalability of spatial color-coded identifiers, we calculate the number of typical $M \times N$ rectangular unique identifiers. In this case, a red row of LEDs with a length of $N$ is considered as an orientation header on one side of the landmark (a typical landmark with $N=3$ is shown in Fig. \ref{fig:System_Setup}). Furthermore, in order to provide white color illumination, an equivalent number of each color, namely, $\frac{M \times N}{3}$ number of red, $\frac{M \times N }{3}$ number of green, and $\frac{M \times N }{3}$ number of blue should be used in the landmarks. The number of aforementioned unique rectangular identifiers, $ {\rm .R.I.}$, is equal to the number of unique identifier in the remaining $(M-1) \times N$ code's points. The number of unique rectangular identifiers can be written as:
\begin{align}
 {\rm R.I.}=&\frac{((M-1)N)!}{ (\frac{(M-3) N }{3})!(\frac{M \times N }{3})!(\frac{M \times N }{3})! }\nonumber
\\ &- \frac{((M-2)N)! S(M-6)}{ (\frac{(M-6) N }{3})!(\frac{M \times N }{3})!(\frac{M \times N }{3})! },
\label{Globalequ1}
\end{align}
where the first term is the number of permutations of the multiset $\{\frac{(M-3) N }{3} {\rm reds}, \frac{M \times N }{3} {\rm greens}, \frac{M \times N }{3}  {\rm blues}\} $ in the rest of $(M-1) \times N$ code's points. The latter term is the number of $M \times N$ rectangular codes with the red row in both sides that are not allowed to be used based on orientation constraints. Similarly, this term is equal to the number of permutations of the multiset $\{\frac{(M-6) N }{3}  {\rm reds}, \frac{M \times N }{3}  {\rm greens}, \frac{M \times N }{3}  {\rm blues}\} $. Moreover, $S(n)$ is the discrete unit step function defined as $S(n\geq 0)=1$ and $S(n<0)=0$.
One can calculate that for a typical $ 6 \times 3 $ landmark there are $419496$ unique identifiers. Hence, this system is significantly scalable in comparison to the other positioning systems with a limited number of identifiers.

\section{Experimental Results and Discussion}
\label{sec:Experimental results and discussion}
\subsection{Experimental Setup}
The prototype system of this proposed indoor positioning algorithm is implemented using a setup of LED luminaries and a typical smartphone. The luminary setup consists of four colors LED landmarks assembled in a cubicle $5\times 5\times 3$ \si{m^3} in size. Given that the proposed algorithm does not require any LED modulation, unmodified commercial LED luminaries could be used. However, LED drivers were designed with modulation capability for developing a hybrid communication and positioning system. 
\begin{figure}[!t]
	\centering
	\includegraphics[width=3.2in]{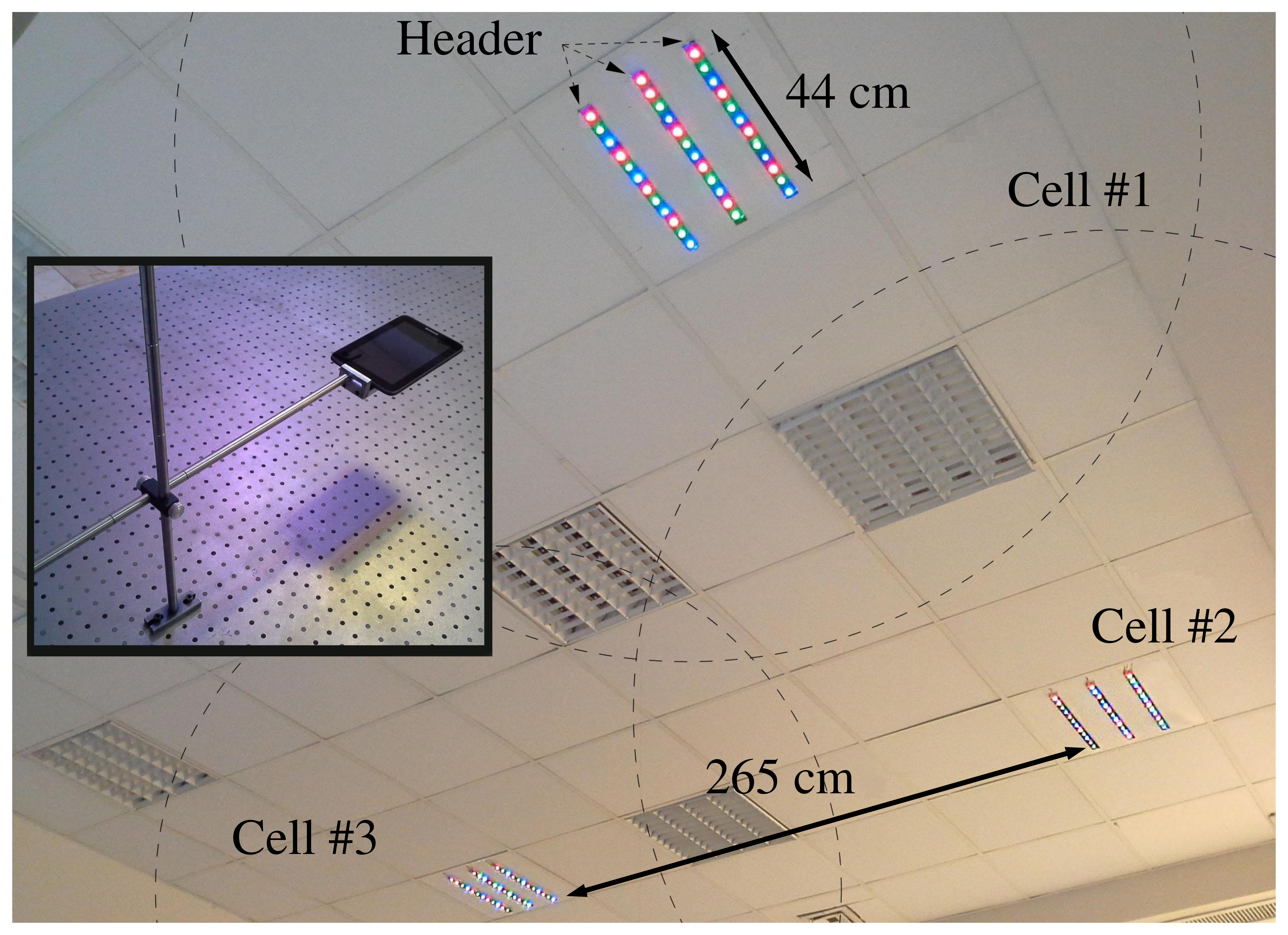} 
	\caption{Indoor positioning prototype in which 4 LED landmarks are installed on the ceiling with
		$300$ \si{cm} height.
	}
	\label{fig:System_Setup}
\end{figure}
Fig. \ref{fig:System_Setup} illustrates the indoor positioning prototype. A typical smartphone is used for capturing the images, measuring the applied tilt, image processing, and positioning algorithm calculations. Taking advantage of an ordinary front-facing camera with a resolution of $640 \times 480$ pixels and an accelerometer, which are available in smartphones, justify the compatibility of our system for commercial applications.
\subsection{Calibration}
\begin{figure}[!t]
	\centering
	\includegraphics[width=3.4in]{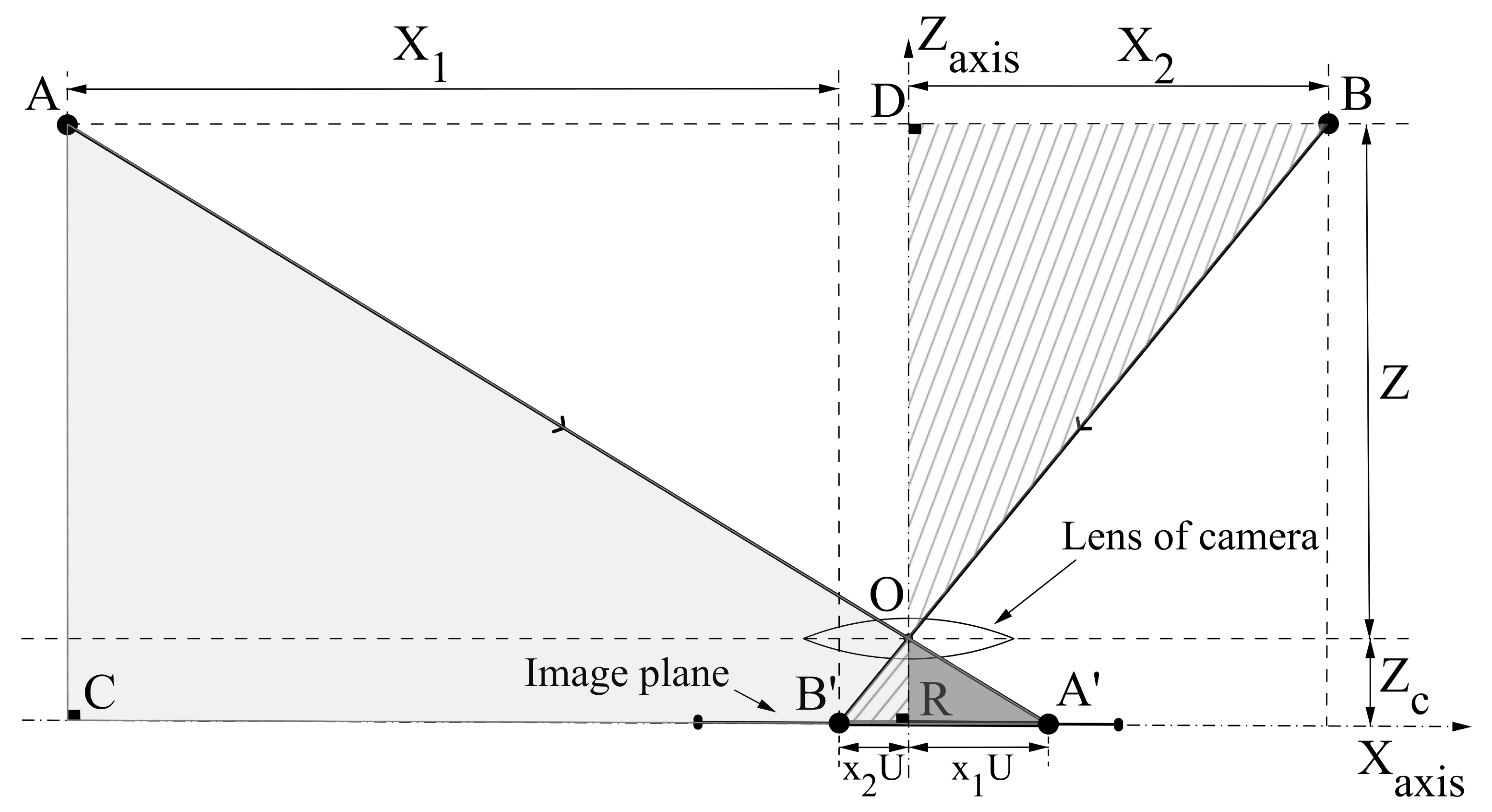} 
	\caption{Initial experiment required for calculating parameters of the camera.}
	\label{fig:3D geometry Scheme}
\end{figure}
In this section, we discuss initial calibration steps for the GOPA system. In order to employ the GOPA algorithm on different smartphones, the camera's parameters of each smartphone, namely $U$ and $Z_{c} $ should be calculated once. In this regard, the camera has to be held in a specific location with known $Z$ and $X$. Fig. \ref{fig:3D geometry Scheme} illustrates the calibration setup, where $A$ and $B$ are the extracted point sources located on the ceiling with the same height, $Z$. $A'$ and $B'$ are the projected images of $A$ and $B$ on the camera's image plane. Considering the similarity of $\triangle\overline{ACA'}$ and $\triangle\overline{ORA'}$, the following holds
\begin{align}  
\frac{Z}{Z_{c} } &=\frac{X_{1} +x_{2} U}{x_{1} U}.
\label{Calequ1}
\end{align} 
In the same way, $\triangle\overline{BDO}$ and $\triangle\overline{ORB'}$ are similar, and thus
\begin{equation}
\frac{Z}{Z_{c} } =\frac{X_{2} }{x_{2} U}.
\label{Calequ3}   
\end{equation}
According to (\ref{Calequ1}) and (\ref{Calequ3}), $U$ and $Z_{c} $ can be calculated as
\begin{align} \label{Calequ4} 
U=\frac{X_{2} x_{1} -X_{1} x_{2} }{x_{2}^{2} },  \nonumber \\
Z_{c} =Z\frac{X_{2} x_{1} -X_{1} x_{2} }{X_{2} x_{2} },
\end{align} 
where $x_{1} $ and $x_{2} $ are measured in the captured image in \si{pixels}. 

\subsection{Experimental Results}
\begin{figure}[!t]
	\centering
	\includegraphics[width=3.4in]{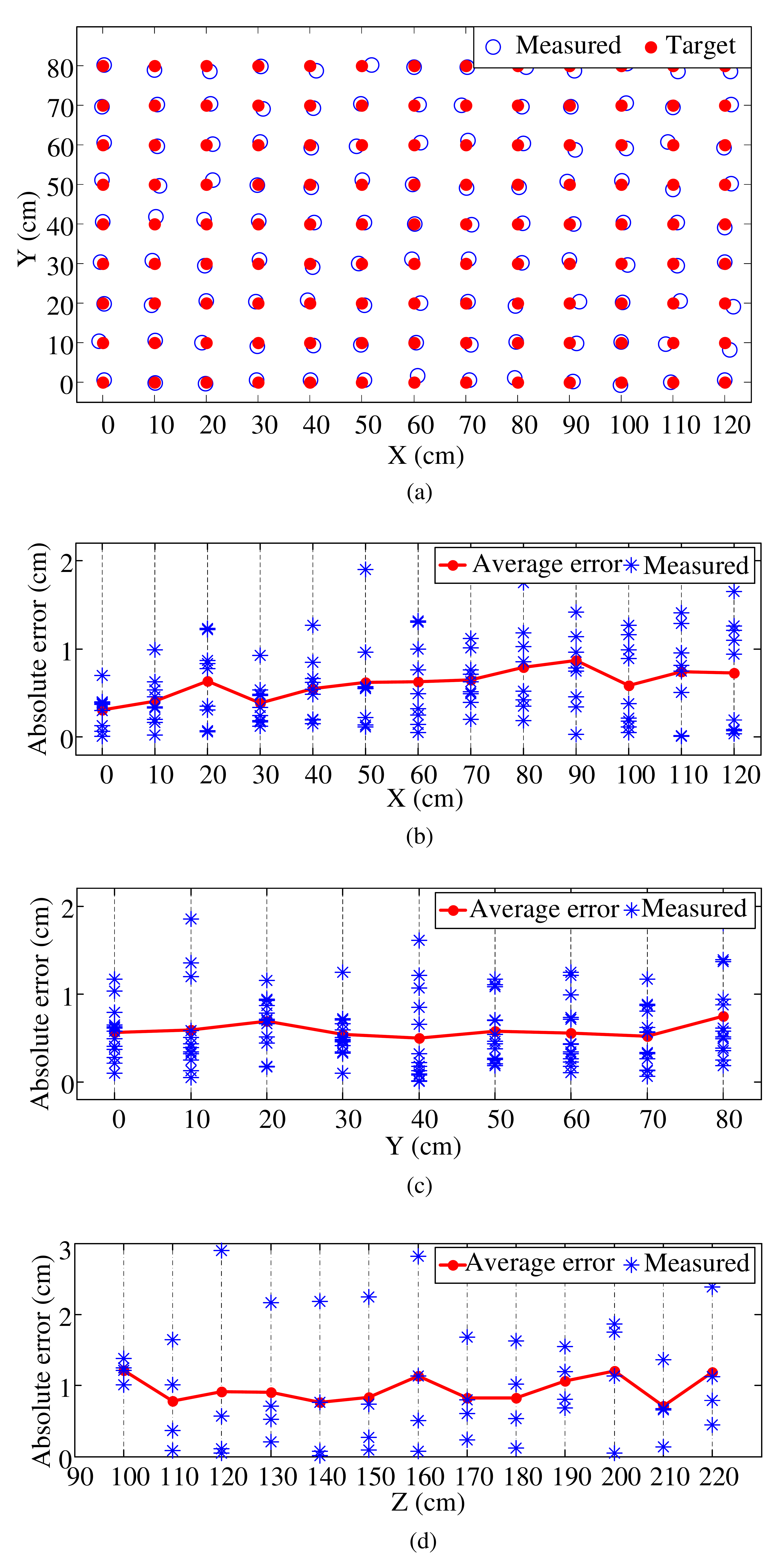} 
	\caption{$xy$ plane measurements. a) $xy$ measurement patterns. b) Position errors at locations $(X,Y, 220)$. The measured $(X, Y)$ coordinates are averaged over different $Y$ values in order to illustrate the algorithm sensitivity to $X$ values. c) Position errors at locations $(X,Y, 220)$. The measured $(X, Y)$ coordinates are averaged over different $X$ values in order to illustrate the algorithm sensitivity to $Y$ values. d) Position errors at locations $(X, Y, Z)$, where $X$ and $Y$ belong to the perimeter of a circle with $1$ meter radius. The measured $(X, Y, Z)$ coordinates are averaged over different heights to illustrate the algorithm sensitivity to $Z$ values.}
	\label{subfigure4}
\end{figure}
In this part, we evaluate the positioning accuracy of GOPA based on experimental measurements. The desired positioning variables that should be evaluated are $X, Y, Z$, azimuth, roll, and pitch. In all of the results, the position error is defined as the Euclidean distance between the real position and its corresponding measured one. Fig. \ref{subfigure4} illustrates the measurement results of the $2$-D positioning algorithm.
The local positioning measurements are done in a $120\times 80$ \si{cm^2} area with a vertical distance of  $220$ \si{cm} below a landmark, as it is shown in Fig. \ref{subfigure4}-(a). The camera has $90^{\circ} $ azimuth and zero tilt.
  Figs. \ref{subfigure4}(b) and \ref{subfigure4}(c) present the $2$-D positioning error as a function of $X$ and $Y$, respectively. The average $2$-D positioning error in a positioning cell zone is $0.54$ \si{cm}. 

Given that $2$-D and $3$-D positioning algorithms have the same calculations for $X$ and $Y$ parameters, the sensitivity of the $3$-D positioning algorithm to $X$ and $Y$ parameters is similar to the $2$-D positioning algorithm. In other words, both the $2$-D and $3$-D positioning errors are almost constant for different horizontal distances from the landmark. In order to evaluate the vertical distance parameter, namely $Z$, we measure the position error on the perimeter of the circle with $1$ meter radius. In addition, we consider different heights ranging from $110$ \si{cm} to $220$ \si{cm} with respect to the ceiling. Fig. \ref{subfigure4}-(d) depicts the $3$-D positioning error versus $Z$. Similarly, the mean $3$-D positioning error remains almost constant for different values of $Z$. 
\begin{figure}[!t]
	\centering
	\includegraphics[width=3.3in]{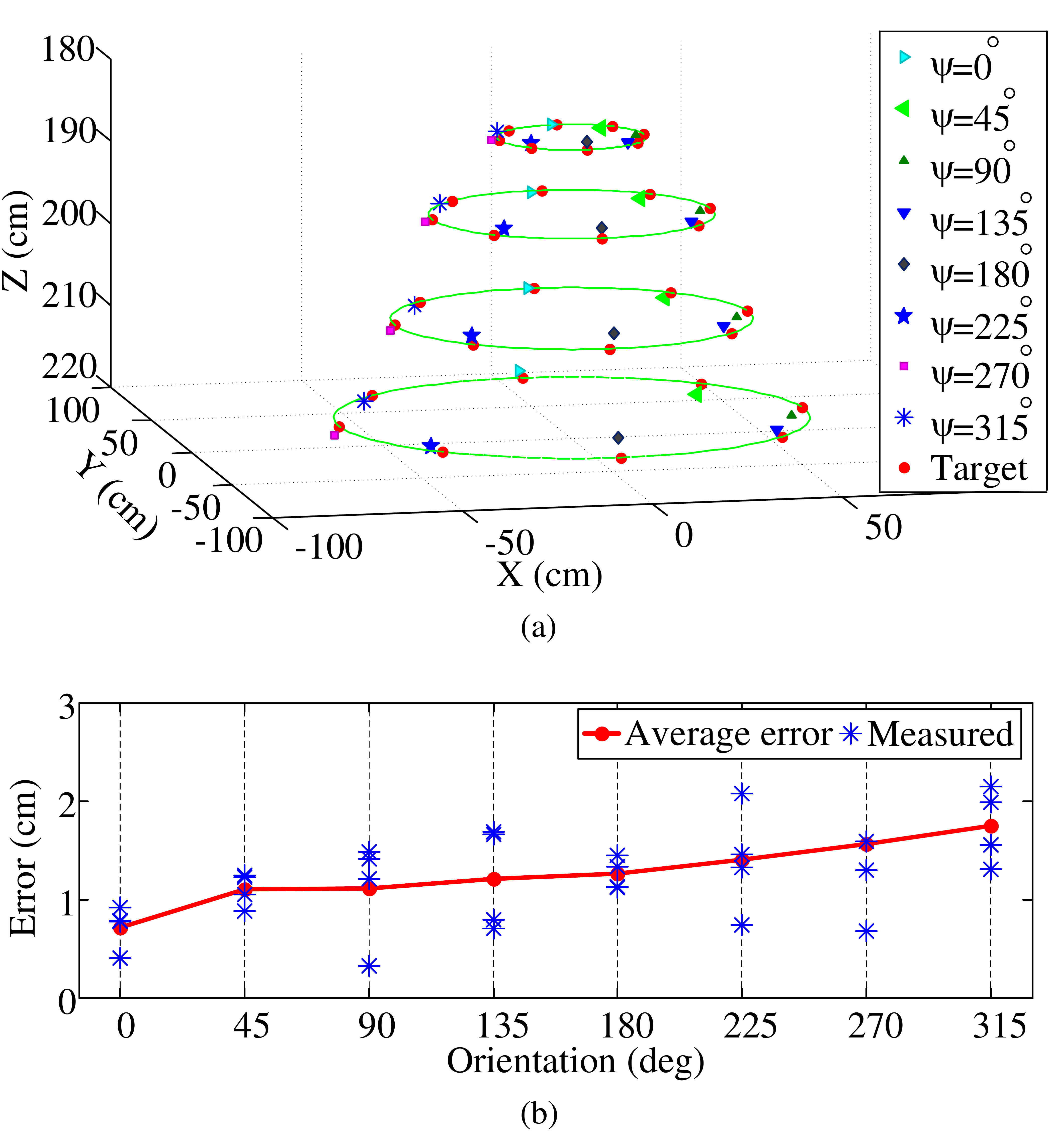} 
	\caption{Demonstration of the experiment to evaluate the system performance caused by azimuth. (a) $3$-D demonstration. (b) Average $3$-D error of the system per different values of orientation.}
	\label{subfigure5}
\end{figure}

Fig. \ref{subfigure5} demonstrates the performance of GOPA for different values of azimuth. In this experiment, measurements are taken for $8$ azimuth angles on the perimeter of the circles for $4$ different heights (see Fig. \ref{subfigure5}-(a)). The mean positioning error increases slightly by the azimuth enhancement as it is shown in Fig. \ref{subfigure5}-(b). This is because the azimuth mapping error grows for higher azimuth degrees.

\begin{figure}[!t]
	\centering
	\includegraphics[width=3.2in]{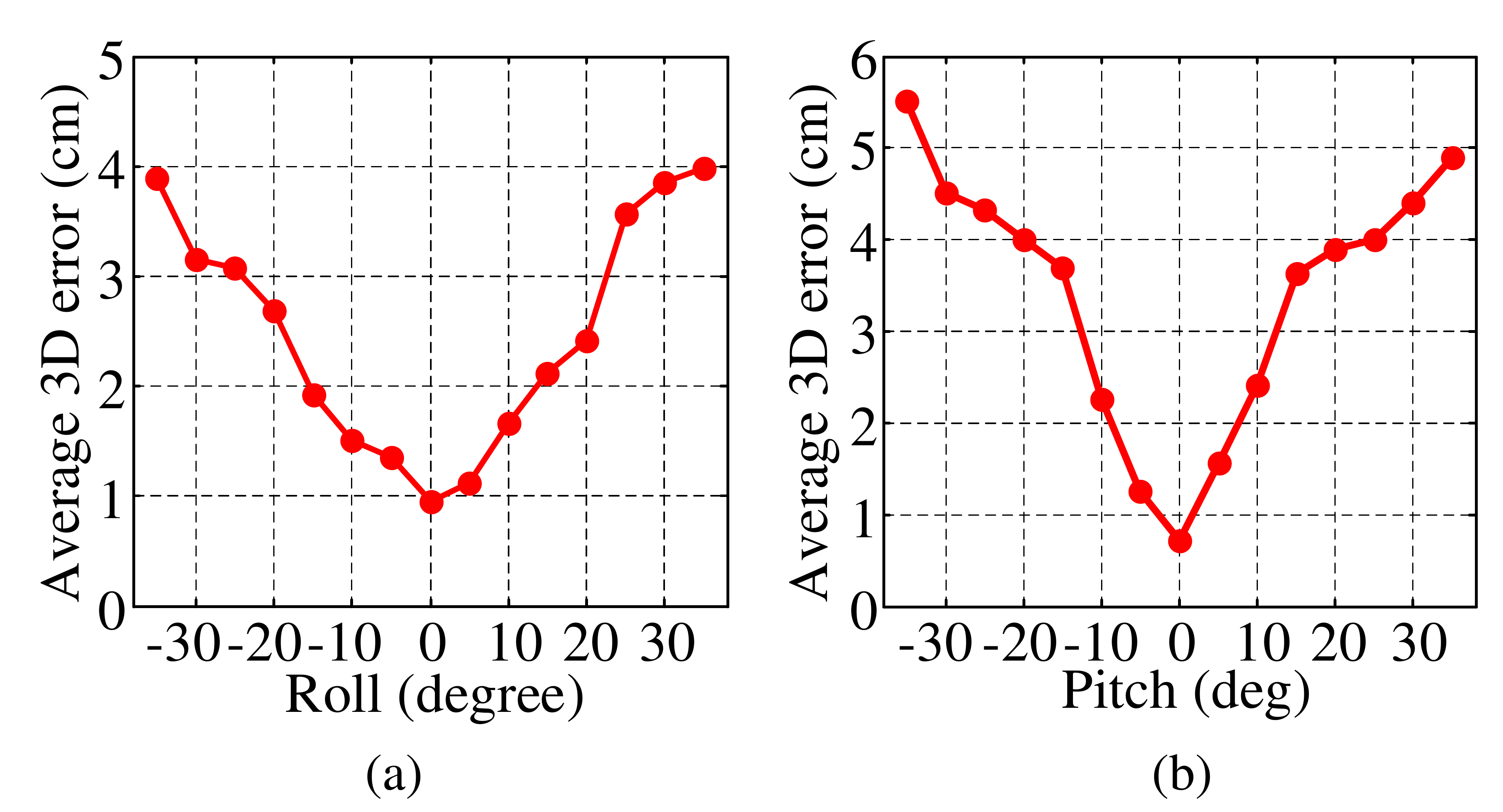} 
	\caption{Algorithm sensitivity to different values of (a) Roll (b) Pitch.}
	\label{subfigure6}
\end{figure}
The results in Figs. \ref{subfigure6}(a) and \ref{subfigure6}(b) are obtained for different values of roll and pitch, respectively. One can conclude that the average positioning error is increased by increasing roll and pitch due to the tilt mapping error. This occurs because the image plane resolution is confined by the size of pixels. On the other hand, the larger the applied tilt to the image plane, the farther the virtual projected point, $P^{''}$, is located (see Fig. \ref{subfigure6}-(b)), i.e., the longer tilt mapping vector. Accordingly, a pixel-level displacement error on the tilted image plane brings about a large displacement error in the virtual projected point proportional to the tilt value. In practice, the applied tilt of the smartphone would not go beyond $35^{\circ}$, while the user is looking at the display. Hence, in the worst case, the average positioning error is less than $6.02$ \si{cm}.

\begin{figure}[!t]
	\centering
	\includegraphics[width=3.3in]{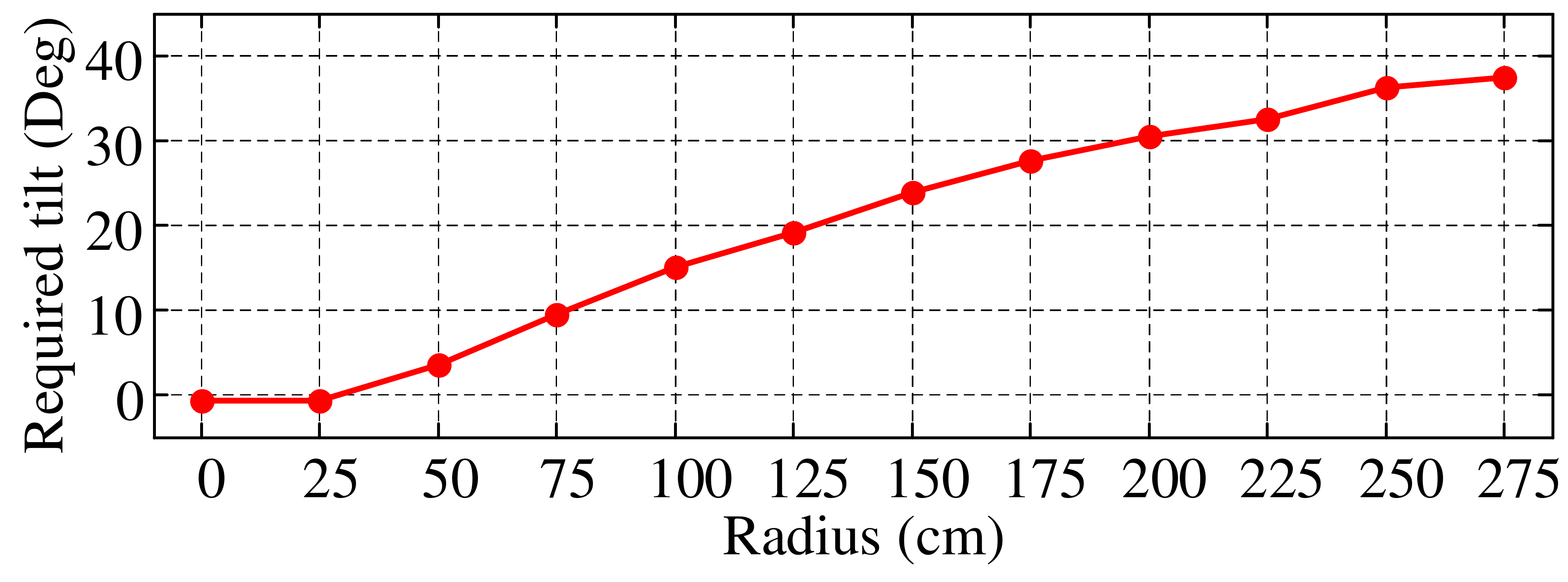} 
	\caption{Required tilt of the smartphone per radius from the landmark center for addressing the camera's FOV problem.}
	\label{fig:Virtual_Plane_M}
\end{figure}

 Another benefit of considering tilt is the enhancement of the FOV of the camera that leads to a larger positioning cell area. Fig. \ref{fig:Virtual_Plane_M} illustrates the required tilt versus the positioning cell radius that is measured in $220$ \si{cm} vertical distance from the ceiling. The cell radius can be increased up to $275$ \si{cm}, applying $35^{\circ}$ tilt to the camera. In this case, $3$-D positioning error is less than $6.02$ \si{cm} (see Fig. (\ref{subfigure6})). Table \ref{table_2} summarizes the performance and also the potential applications of GOPA.

\begin{table}[t]
	\begin{center}
		\caption{System performance and applications.}
		\label{table_2}
		\begin{tabular}{|c|c|c|c|c|}
			\hline
			\textbf{\!Dimension\!}            & \textbf{\begin{tabular}[c]{@{}c@{}}\!Azimuth\! \\ (deg.)\end{tabular}} & \textbf{\begin{tabular}[c]{@{}c@{}}\!Tilt\! \\ (deg.)\end{tabular}} & \textbf{\begin{tabular}[c]{@{}c@{}}\!Average\!\\ Error\\  (cm)\end{tabular}} & \textbf{\!Potential\! Applications}                                                                                         \\ \hline \hline
			\textbf{$2$-D}                  & $90$ (fixed)                                                          & $0$                                                               & $0.54$                                                           & \begin{tabular}[c]{@{}c@{}}Hospital wheelchairs\\and robots\end{tabular}                                   \\ \hline
			\multirow{3}{*}{\textbf{$3$-D}} & $0$                                                                  & $0$                                                               & $1.24$                                                           & \multirow{3}{*}{\begin{tabular}[c]{@{}c@{}}Realistic positioning \\ applications on\\ smartphones \end{tabular}} \\ \cline{2-4}
			& $275$                                                                & $0$                                                               & $1.85$                                                           &                                                                                                              \\ \cline{2-4}
			& $275$                                                                & $35$                                                              & $6.02$                                                           &                                                                                                              \\ \hline
		\end{tabular}
	\end{center}
\end{table}
\section{Conclusion}
In this paper, we present a practical and low complexity visible light indoor positioning system on an unmodified smartphone using spatial color-coded landmarks and geometrical optics algorithms. The proposed system utilizes the front-facing camera and the built-in accelerometer of the smartphone to compute the distance from the landmark. The introduced virtual plane approach addresses the tilt problem of the camera, as well as the FOV constraint that is reported in most AOA-based positioning systems. Experimental results show an average positioning error of less than $0.54$ \si{cm} and $1.24$ \si{cm} for no tilt 2-D and 3-D positioning, respectively. Applying the $35^{\circ}$ tilt to the camera as the worst condition in practice, the average positioning error remains at less than $6.02$ \si{cm}.
\section{Acknowledgement}
Parts of this work including the idea and experimental results have been protected by us patent US9939275B1 \cite{salehi2018methods}. 
\ifCLASSOPTIONcaptionsoff
\newpage
\fi
\bibliographystyle{IEEEtran}
\bibliography{VLP}

\end{document}